\documentstyle[12pt]{article}

\advance\textheight by -5cm 
\advance\textheight by -2.5cm 
\begin{document}
\begin{titlepage}
July, 1995          \hfill
    LBL-37524 \\
\mbox{} \hfill    UCB-PTH-95/26
\begin{center}
{\large \bf General Classical Solutions of Nonlinear\\
 $\sigma$-Model and Pion
Charge Distribution  \\
of Disoriented Chiral
Condensate\footnote{This work was
supported in part by the Director, Office of
Energy Research, Office of High Energy and Nuclear Physics, Division of
High Energy Physics of the U.S. Department of Energy under Contract
DE-AC03-76SF00098 and in part by the National Science Foundation under
Grant PHY-90-21139.\\
$^\ddagger$ Also at Department of Physics, University of California,
Berkeley, CA 94720.}}
\vskip 0.5cm
 Zheng Huang$^\dagger$ and Mahiko Suzuki$^\ddagger$

\vskip 0.3cm
$^\dagger${\em Department of Physics, University of Arizona, Tucson,
AZ 85741}\\
$^\ddagger${\em Theoretical Physics Group, Lawrence Berkeley Laboratory\\
University of California,       Berkeley, CA 94720, USA}
\end{center}

\vskip .2in
\begin{abstract}
We obtain  the general analytic solutions of the nonlinear $\sigma$-model
in $3+1$ dimensions as the
candidates for the disoriented chiral condensate (DCC).
The nonuniformly isospin-orientated solutions are shown to be related to
the uniformly oriented ones through the chiral
(axial) rotations. We discuss the pion charge distribution arising from these
solutions. The
distribution $dP/df=1/(2\sqrt{f})$ holds for the uniform solutions
in general and the nonuniform solutions in the $1+1$ boost invariant case.
For the nonuniform solution in $1+1$ without a boost-invariance and in
higher dimensions, the distribution does not hold in the integrated form.
However, it is applicable to the pions
selected from a small segment in the momentum phase space. We suggest that
the nonuniform DCC's may correspond to the mini-Centauro events.
\end{abstract}
\end{titlepage}
\renewcommand{\thepage}{\roman{page}}
\setcounter{page}{2}






\newpage
\renewcommand{\thepage}{\arabic{page}}
\setcounter{page}{1}
\section{Introduction}
The soft particle production in a very high energy hadron-hadron
or nucleus-nucleus collision is an interesting phenomenon.
Occasionally, the collision creates a large number of low
energy (small $p_t$) particles, mainly pion quanta, initially
populating in a small interaction volume and subsequently undergoing
a rapid expansion. The perturbative QCD is not applicable
in describing the dynamics since it involves a large number of
quanta  and the interactions are highly nonlinear. One may
anticipate some novel dynamical feature of the nonperturbative
QCD. Although there is some evidence that the $p_t$ distribution
of these particles follows the scaling law with an effective
temperature, it is not clear whether or not these low energy
particles can actually thermalize so that their distribution can
be described by thermodynamics. In fact, some deviation from
the thermal distribution in the very small $p_t$ region, say,
$p_t<100$ MeV, has been observed though data are poor in this
region at present.

On the other hand, it has been suggested, first by
Horn and Silver \cite{horn}, that these low energy
pions may be described by a classical theory. The number ($N$) of
quanta involved is large, the quantum fluctuation is suppressed
by $1/\sqrt{N}$. In addition, the low energy theorem on the
pion-pion scatterings dictates that the quantum corrections to
the scattering amplitudes are suppressed by a factor of
$p^2/(4\pi f_\pi)^2$. More recently, a scenario of disoriented
chiral condensate \cite{DCC,rw} suggests that these
low energy pions may be out of equilibrium and undergo a quench
following a chiral phase transition, and their interactions
should be described by the classical chiral dynamics.

In this paper, we shall determine the possible classical
field evolutions that these low energy pions may follow
based on the nonlinear $\sigma$-model.
The advantage of the {\it nonlinear} $\sigma$-model over
the {\it linear} $\sigma$
model is  that the constraint of the vacuum expectation value on
the fields is
built in and that the pion fields always describe the massless modes
irrespective of the vacuum orientation in the background.
The $\sigma$ mass is taken to be infinity so the low energy
structure of the theory is evident.
We have obtained in an analytic form a  class of classical solutions to
the nonlinear $\sigma$-model in $3+1$ space-time dimensions as the
candidates
of the disoriented chiral condensate in QCD.
Our general solutions have a
transverse momentum distribution and need not
be subject to a boost-invariance
constraint.  The solution with a nonuniform isospin orientation is constructed
by the chiral $SU(2)\times SU(2)$ rotation from a uniformly oriented solution.
In the limit of a boost invariance and no transverse momentum,
our solutions reduce to those of Blaizot and Krzywicki \cite{blaiz}.
We study the distribution of the neutral pion fraction $f$ for the pions
that disintegrate from the disoriented vacua.  We find that the
distribution dP/df = 1/(2$\sqrt{f})$ holds for the uniformly oriented vacua
and also for the boost-invariant vacua with an infinitely large uniform spread
in the transverse direction, but it does not hold for the vacuum whose
isospin orientation is nonuniform in space-time. However this distribution
should be correct if one selects pions from within a small region in the
$y$-${\bf k}_\bot$ plot event by event.  This conclusion is reached
through the analysis using the classical field theory method and also by
studying the  quantum pion states.

  We organize the paper as follows.  In Sec.\ 2, we start with the
analysis for
the boost-invariant solutions with no transverse momentum.  In Sec.\ 3, we
make the observation that all solutions with a nonuniform isospin orientation
are obtained by the chiral rotations from
 a uniformly oriented solution whose energy
is degenerate with the nonuniform ones.  In Sec.\ 4, we give a general
solution with a uniform isospin orientation, from which we can
obtain the nonuniform solutions by the chiral rotations according to
the prescription given in Sec.\ 3.  The general solution has a nontrivial
transverse momentum distribution and is not subject to
the  boost-invariance constraint.
In Sec.\ 5, we examine the charge distribution of the pions
disintegrating from these disoriented vacua.  The picture of classical field
theory leading to the distribution dP/df = 1/(2$\sqrt{f})$ does not apply to
the nonuniformly oriented vacua except in the boost invariant limit with zero
transverse momentum.  For a general
solution which has rapidity and transverse momentum dependence,
the distribution  holds only within
each small segment in the $y$-${\bf k}_\bot$ plot.
The charge distribution is
also studied  from the viewpoint of quantum multipion states, following
Horn and Silver \cite{horn}.  The modification of the
distribution  is attributed to the fact that
for the general nonuniform solution, more than one orbital state
is available
for pions to occupy so that there are many different ways to construct
multipion states.

\section{Boost-Invariant Solution in $1+1$ Dimensions}
In high energy hadron or nucleus collisions, the
configurations approximately invariant
along the collision axis are of particular interest.
We first focus on this
class of solutions ignoring the transverse spatial dependence.
We choose a {\it nonlinear} $\sigma$-model
as the dynamical model for QCD at low energy.
  The solutions that
we obtain in this Section are equivalent to those of Blaizot and Krzywicki
$\cite{blaiz}$ though they are dervied in a slightly different way in order
to clarify a relation between the
uniformly oriented solutions and the nonuniformly
oriented ones which  plays an important  role
when we extend our argument to the more general case later.

The phase and radial
representation of the nonlinear $\sigma$-model is,
\begin{equation}
      \Sigma (x) = e^{i{\mbox{\boldmath $\tau$}
\cdot{\bf n}(x)\theta(x)}}.
\end{equation}
No matter what values the classical phase fields take, the state remains at the
bottom of the potential valley because $|\Sigma| = 1$.  This facilitates
greatly the search for the DCC-type solutions which are realized at the
bottom of potential well.  Define the pion field
\begin{equation}
    \mbox{\boldmath $\pi$}
 (x) = f_{\pi}{\bf n}(x)\,\theta(x).
\end{equation}
where ${\bf n}(x)$ is an unit
isovector field obeying ${\bf n}(x)\cdot{\bf n}
(x) = 1$.  Alternatively one defines $\mbox{\boldmath $\pi$} (x)$
by $\sigma + i\mbox{\boldmath $\tau\cdot\pi$} =
f_{\pi}\Sigma$ with the constraint $\sigma =\sqrt{f_{\pi}^2 -
\mbox{\boldmath $\pi$}^2}$.
In this case the pion fields are given by $\mbox{\boldmath $\pi$}(x) =
f_{\pi}{\bf n}(x)\:{\rm sin}\theta(x)$. In either case,
${\bf n}$ determines the isospin orientation  of the pion field.

The  lagrangian is given by
\begin{equation}
{\cal  L} = \frac{f_{\pi}^2}{4}
{\rm tr} \Bigl(\partial_{\mu}\Sigma^\dagger(x)
    \partial^{\mu}\Sigma(x)\Bigr), \nonumber
\end{equation}
where $\Sigma$ transforms like $\Sigma\rightarrow U_L\Sigma U_R^\dagger$ under
$SU(2)_L\times SU(2)_R$ rotations.
   In terms of $\theta(x)$ and ${\bf n}(x)$,
the lagrangian is
\begin{equation}
  {\cal L} = \frac{f_{\pi}^2}{2}(\partial_{\mu}\theta\:\partial^{\mu}\theta +
     \sin^{2}\theta\partial_{\mu}{\bf n}\cdot\partial^{\mu}{\bf n}) +
     \frac{\lambda f_{\pi}^2}{2}({\bf n}^2 -1),
\end{equation}
where $\lambda$ is a Lagrange multiplier.
We will not include an explicit chiral symmetry breaking throughout this paper.
The Euler-Lagrange equations are
\begin{equation}
\Box\theta = \sin\theta\:
\cos\theta\:\partial_{\mu}{\bf n}\cdot\partial^{\mu}
     {\bf n},
\end{equation}
\begin{equation}
     \partial_{\mu}(\sin^{2}\theta\:\partial^{\mu}{\bf n}) = \lambda{\bf n}.
\end{equation}
The chiral $SU(2)_{L}\times SU(2)_{R}$ symmetry assures the
conservation of the
vector and axial-vector currents.  In terms of $\theta$ and ${\bf n}$,
the current conservation is written as
\begin{equation}
   \partial_{\mu}(\sin^{2}\theta\:{\bf n}\times\partial^{\mu}{\bf n}) = 0,
\end{equation}
\begin{equation}
   \partial_{\mu}({\bf n}\:\partial^{\mu}\theta + \sin\theta\:
   \cos\theta\:\partial^{\mu}{\bf n}) = 0.
\end{equation}
The isospin current conservation  (7) follows also from  (6), while
the axial-vector current conservation  (8) can be derived from
 (5) by repeated use of $({\bf n}\cdot d{\bf n}/d\tau) = 0$.
 (4), (5), (6) and (7) are most general with no assumptions or
approximations made.

We consider a boost-invariant case in 1+1 dimensions where
the fields $\theta(x)$ and ${\bf n}(x)$ are only functions  of the variable
$\tau$:
\begin{equation}
            \tau = \sqrt{t^{2}-x^{2}}.
\end{equation}
For a function only of $\tau$, a partial derivative
$\partial_{\mu}f(\tau)$ is equal to $(x_{\mu}/\tau)df/d\tau$.
Furthermore, $\partial_{\mu}(f(\tau)\partial^{\mu}g(\tau)) =
(1/\tau^2)(d(\tau^{2}fg^\prime)/d\tau)$ where $g^\prime = dg/d\tau$.
The current conservation relations can be integrated into
\begin{equation}
   \tau \: \sin^{2}\theta\:{\bf n}\times\frac{d{\bf n}}{d\tau}= {\bf a},
\end{equation}
\begin{equation}
 \tau\,{\bf n}\frac{d\theta}{d\tau}+\tau\,\sin\theta\:\cos\theta\:
\frac{d{\bf n}}{d\tau}= {\bf b},
\end{equation}
where ${\bf a}$ and ${\bf b}$ are constant vectors in the isospin space
whose magnitudes are denoted as $a$ and $b$ respectively.
It is immediately obvious from  (10) and (11) that ${\bf a}$
and ${\bf b}$ are orthogonal to each other:
\begin{equation}
                {\bf a}\;\bot\;{\bf b}.
\end{equation}
The isovector field ${\bf n}(\tau)$ stays perpendicular to ${\bf a}$
as $\tau$ varies.  By multiplying  (10) with ${\bf n}$ vectorially and using
${\bf n}\cdot (d{\bf n}/d\tau) = 0$, one obtains
\begin{equation}
   \frac{d{\bf n}}{d\tau} = \frac{{\bf a}\times{\bf n}}{\tau\sin^{2}\theta},
\end{equation}
a standard equation for a vector ${\bf n}$
 to precess around a constant vector ${\bf a}$. The
precession frequency  $|{\bf a}|/\tau\sin^2\theta$
varies with the proper time $\tau$. The relations among ${\bf a}$,
${\bf b}$ and ${\bf n}$ are illustrated in Figure 1.
Squaring (11) gives
\begin{equation}
 \Biggl(\tau\frac{d\theta}{d\tau}\Biggr)^{2}+ \sin^{2}\theta\:\cos^{2}\theta\
\Biggl(\tau\frac{d{\bf n}}{d\tau}\Biggr)^{2} = b^{2}.
\end{equation}
Eliminating $d{\bf n}/d\tau$ from the these equations, one obtains the
differential equation for $\theta(\tau)$:
\begin{equation}
    \Biggl(\tau\frac{d\theta}{d\tau}\Biggr)^{2} = a^{2}+ b^{2} -
    \frac{a^{2}}{\sin^{2}\theta},
\end{equation}
where $a = |{\bf a}|$ and $b = |{\bf b}|$.
 (13) and (15) combined contain the same information as the first integrals
of the Euler-Lagrange equations for $\theta$ and ${\bf n}$ so that
we may proceed with the current conservation laws.

   (15) is analytically integrable into the most general boost-invariant
solution for $\theta(\tau)$ in 1+1 dimension:
\begin{equation}
 \cos\theta(\tau) = (b/\kappa)\; \cos\Bigl(\kappa\:{\rm ln}(\tau/\tau_0)
+ \vartheta_0\Bigr),
\end{equation}
where $\kappa^2 = a^2+b^2$ and $ \cos\vartheta_0 =(\kappa/b)\,\cos
\theta
(\tau_0)$.
Substituting $\theta(\tau)$ in the integral from the isovector current
conservation, one obtains the general solution for ${\bf n}(\tau)$. Since
${\bf n}(\tau)$
and ${\bf b}$ both lie in the plane perpendicular to ${\bf a}$, it is
convenient to express the unit isovector field ${\bf n}(\tau)$ in terms of
a single angle $\beta(\tau)$ measured from the direction of ${\bf b}$:
\begin{equation}
      {\bf n}(x)\cdot{\bf b} = b \cos\beta(x).
\end{equation}
The solution for $\cos\,\beta(\tau)$ can be expressed
in terms of  $\theta(\tau)$
\begin{equation}
 \cos\,\beta(\tau) =\frac{a}{b}\sqrt{\frac{\kappa^2}{a^2}
         -\frac{1}{\sin^{2}\theta(\tau)}}.
\end{equation}
The DCC configuration may be obtained
 from the  boost-invariant solution multiplied by
a step function ${\rm\Theta}(\tau^2)$:
\begin{equation}
   \Sigma(x)_{DCC} = e^{i\theta (x)\mbox{\boldmath $\tau$ }
\cdot{\bf n}(x)}{\Theta}(\tau^2),
\end{equation}
such that the causality condition
 is satisfied.  Our solutions should only apply to the
inside of the light-cone.
Once  $\Theta(\tau^2)$ is inserted,
there appears a source on the light cone at $\tau = 0$ which triggers
the formation of a DCC.  The energy density ${\cal E}(x)$ of our solution is
singular as we approach the light cone:
\begin{equation}
 {\cal  E}(x) = \frac{f_{\pi}^2}{2}
                 \Biggl(\frac{t^2+z^2}{\tau^2}\Biggr)(a^2+b^2),
\end{equation}
for both the uniform and nonuniform solutions.  The isospin vectors ${\bf a}$
and ${\bf b}$ enter the energy density in the combination of $a^2+b^2$ as
required by $SU(2)\times SU(2)$ invariance.  The lowest energy solution of
$a^2+b^2 =0$ is a trivial solution obtained from  (16), (17) and (18)
by taking the limit of $a, b \rightarrow 0$:
\begin{equation}
        \theta(x) = {\rm constant},\; {\bf n}(x) = {\rm constant\;vector}.
\end{equation}
Blaizot and Krzywicki \cite{blaiz} expressed the pion fields
by $\pi = f_{\pi}{\bf n}\,\sin\theta$.  If we choose
our $\it{initial}$ condition such that $\sin\,\vartheta_0 = 0$, that is,
$\cos\theta(\tau_0)=b/\kappa$, our general solution given by  (16) and (18)
coincides with theirs.

\section{Chiral Rotation and Nonuniform Solution}
In this Section we focus on the relation between the
space-time dependence of the isovector field ${\bf n}$(x)
 and $SU(2)\times SU(2)$ invariance of the lagrangian.
It is easy to see in  (18) that our solutions have a uniform isospin
orientation when the isospin vector ${\bf a}$ vanishes. When ${\bf a}
=0$, $\beta(\tau) = 0\;({\rm mod}\; 2\pi)$, that is, ${\bf n}$ points to the
direction of ${\bf b}$ for all uniform DCC's.  Because of the
$SU(2)\times SU(2)$ invariance, it is always possible to rotate the vector
${\bf a}$ by an appropriate axial rotation to the direction of the vector
${\bf b}$. After the rotation the solution has a uniform isospin
orientation and is degenerate in energy with the nonuniform solution
prior to the rotation.  It is not unfamiliar that if a system possesses
some
symmetry,  a set of infinitely many new solutions may be obtained by making
the symmetry transformation on a single solution.

 To be explicit in the present case, let us rotate a uniform solution
\begin{equation}
\Sigma_0 (x) = e^{i\theta_0(x)\mbox{\boldmath $\tau$}\cdot{\bf n}_0},
\end{equation}
where ${\bf n}_0$ is space-time independent.  Upon a
general chiral
rotation  parametrized by $U_L = e^{i\xi\mbox{\boldmath $\tau$}
\cdot{\bf n}_L}$ and $U_R =
e^{i\eta\mbox{\boldmath $\tau$}
\cdot{\bf n}_R}$, the uniform solution is rotated into $\Sigma (x)
= U_L\Sigma_0 (x) U_R^{-1}$.  The transformed $\theta$ and ${\bf n}$
fields are given by
\begin{eqnarray}
c_\theta & = & \Bigl(c_\xi c_\eta + ({\bf n}_L\cdot{\bf n}_R)s_\xi s_\eta
               \Bigr) c_0 + \nonumber \\
         &   & \Bigl(({\bf n}_0\cdot{\bf n}_R)c_\xi s_\eta -
               ({\bf n}_0\times{\bf n}_L) s_\xi c_\eta +({\bf n}_0\times
                {\bf n}_L)\cdot{\bf n}_R\, s_\xi s_\eta\Bigr)s_0,
\end{eqnarray}
and
\begin{eqnarray}
{\bf n} s_\theta  & = &\Bigl({\bf n}_L s_\xi c_\eta -{\bf n}_R c_\xi s_\eta
  +({\bf n}_L\times{\bf n}_R)s_\xi s_\eta\Bigr) c_0 \nonumber \\
                 &   &  \mbox{} + \Biggl({\bf n}_0\, c_\xi c_\eta +
              ({\bf n}_0\times{\bf n}_L)s_\xi c_\eta +
              ({\bf n}_0\times{\bf n}_R)c_\xi s_\eta \nonumber \\
            &   & \mbox{} +\Bigl(({\bf n}_0\cdot{\bf n}_L){\bf n}_R +({\bf n}
                               _0\cdot{\bf n}_R){\bf n}_L
              -({\bf n}_L\cdot{\bf n}_R){\bf n}_0\Bigr)s_\xi s_\eta\Biggr)s_0,
\end{eqnarray}
where $c_{\theta}$ and $s_{\theta}$ stand for $\cos\theta$ and $\sin\theta$,
respectively, and so forth, while $c_0 = \cos\theta_0$ and $s_0 = \sin
\theta_0$.
For an isospin rotation, we choose $\xi = \eta$ and ${\bf n}_L = {\bf n}_R$.
Then the rotated fields are
\begin{eqnarray}
  \cos\theta(\tau)&=& \cos\theta_0(\tau), \nonumber \\
  {\bf n}     & =& ({\bf n}_0\cdot{\bf n}_L){\bf n}_L - \Bigl({\bf n}_0 -
                       ({\bf n}_0\cdot{\bf n}_L){\bf n}_L\Bigr)\cos 2\xi
                        + ({\bf n}_0\times{\bf n}_L)\,\sin 2\xi.
\end{eqnarray}
Since this is a global isospin rotation, the resulting field is another
uniformly oriented solution with the same $\theta (x)$.
For an axial rotation, $\xi = \eta$ and ${\bf n}_L = - {\bf n}_R$,  in
particular, if  ${\bf n}_L, {\bf n}_R \;\bot\;
{\bf n}_0$, one obtains
\begin{eqnarray}
 \cos\theta(\tau) & = & \cos 2\xi\: \cos\theta_0(\tau), \nonumber \\
 {\bf n}(\tau)\,\sin\theta(\tau) & = & {\bf n}_0\, \sin\theta_0(\tau) +
                   {\bf n}_L\, \sin 2\xi\: \cos\theta_0(\tau).
\end{eqnarray}
The axial rotations turn a uniform solution
into nonuniform solutions. We can
actually show that our general nonuniform solution given by  (16) and (18)
is reproduced with a suitable choice of the rotation angle:
\begin{equation}
   \tan 2\xi = a/b.
\end{equation}
Then  (26) gives
\begin{eqnarray}
   \cos\theta(\tau) & = & (b/\kappa)\cos\theta_0(\tau), \nonumber \\
   \cos\,\beta(\tau) & = & {\bf n}\cdot{\bf n}_0
             =  \sin\theta_0(\tau)/\sin\theta(\tau).
\end{eqnarray}
The result in (18) for $\cos\beta$ is obtained by solving
the above equations.
In this way we are able to obtain the
nonuniform solutions from uniform ones by the
axial-vector rotations.
In the boost-invariant 1+1 dimensional case, we have obtained in Sec.\ 2
all nonuniform
solutions by solving explicitly the nonlinear differential equations.
We have shown that they are
all related by the axial rotations to the nonuniform solutions
with the same energy density ${\cal E}(x)$.
All possible solutions are exhausted in this way.

\section{General Solution Without Boost Invariance}
  Even without boost invariance, it is straightforward to solve for
the
uniformly oriented solutions. Once we obtain a uniform solution, we can
transform it into the
nonuniform solutions by $SU(2)\times SU(2)$ rotations. The
equations for $\theta(x)$ and ${\bf n}(x)$ from  (5) to (8) in Sec.\ 2 are
also valid in the boost-noninvariant case. We look for the uniform solution in
which ${\bf n} = {\rm constant}$
so that $\partial_{\mu}{\bf n}=0$. In this case the
equation of motion reduces simply to
\begin{equation}
          \Box\theta = 0.
\end{equation}
It is convenient to use the space-time variables
\begin{equation}
 \tau =\sqrt{t^2 - z^2},\;\;\; \eta =\frac{1}{2}
{\rm ln}\Biggl(\frac{t+z}{t-z}\Biggr),
\;\;\;  {\bf x}_\bot.  \nonumber
\end{equation}
The origin of space-time coordinates is identified with the collision
 point of
the hadron collisions.  The z-axis is chosen along the collision axis of the
initial hadrons.  Note that the meaning of variable $\tau$ is a little
different from the 1+1 dimensional case.
The surface of $\tau=0$ lies outside
the light cone with respect to the space-time origin except for the exactly
forward and
backward directions. With these space-time variables, $\Box\theta = 0$ becomes
\begin{equation}
    \frac{1}{\tau}\frac{\partial}{\partial\tau}\Biggl(\tau
\frac{\partial\theta}
{\partial\tau}\Biggr) -
\frac{1}{\tau^2}\frac{\partial^2\theta}{\partial^2\eta}-
 \triangle_\bot\theta = 0.
\end{equation}
Since the differential equation is homogeneous, it can be solved by
the method of separation of variables in the form
\begin{equation}
 \theta(x) = T(\tau)H(\eta)X({\bf x_\bot})\:{\rm\theta}(\tau^2-{\bf x}_\bot^2).
\end{equation}
We solve for $\theta(x)$ inside the light cone, $\tau^2 - {\bf x}_\bot^2 > 0$.

     For the transverse direction
${\bf x_\bot}$, the general solutions are the Bessel and the Neumann
functions.  If we require that the solution be regular on the collision axis
$\rho =|{\bf x_\bot}|= 0$, the Neumann functions are excluded.  Note however
that a singular behavior means an infinite oscillation toward $\rho = 0$, not
an indefinite increase, in terms of the pion field
$\mbox{\boldmath $\pi$} = f_{\pi}{\bf n}\,
\sin\theta$.  One may also require that
$X({\bf x_\bot})$ should not increase
indefinitely as $\rho \rightarrow\infty$.
  With these requirements, the  parameter $\mu^2$
defined by
$\triangle_\bot X = -\mu^2 X$ must be positive.  We choose $\mu
> 0$. $X({\bf x}_\bot)$ is expressed in terms of
 the Bessel functions of integer
order:
\begin{equation}
 X({\bf x_\bot})  =  C_0J_0(\mu\rho)
    + \sum_{m=1}^\infty J_m(\mu\rho)(C_m \cos m\phi + D_m \sin m\phi),
\end{equation}
where $C_0$, $C_m$, and $D_m$ are the numerical
coefficients to be determined by the boundary conditions. The
magnitude of $\mu$ determines a transverse size of a DCC and
therefore a spread of the $p_t$ distribution
of the final pions.  Since a
DCC will have an extended size in the transverse direction, the value of $\mu$
is likely to be a fraction of $f_{\pi}$ or less.

    The spatial rapidity dependence $H(\eta)$ is simply solved
\begin{equation}
  H(\eta) =  \cosh \lambda\eta \;\;\; or \;\;\;\sinh \lambda\eta.
\end{equation}
The parameter $\lambda$ can be any complex number in general (If it is complex,
one should take the real part of $\theta(x)$ at the very end).
  The special case
$\lambda = 0$ leads to the boost-invariant solutions.
For the approximately boost-invariant DCC configurations, the magnitude of
$\lambda$ is much smaller than unity.  The region of large values of $\eta$
corresponds to the forward and backward edges of DCC where energetic leading
hadrons are moving outward while the small values of $\eta$ describe the cool
central region.  In this picture, it appears appropriate to choose
$H(\eta)$ such that the pion density is higher at a larger
 $\eta$ than at a smaller
$\eta$.  We therefore choose $\lambda$ to be real (and positive) in the
following.  It should be emphasized however that our choice
for a  real $\lambda$
over purely imaginary or complex $\lambda$ is more for the convenience
of the presentation.

Given $X({\bf x_\bot})$ and $H(\eta)$, $T(\tau)$ obeys the
differential equation
\begin{equation}
 \frac{1}{\tau}\frac{\partial}{\partial\tau}\Biggl(\tau\frac{\partial T}
{\partial \tau}\Biggr) + \Biggl(\mu^2 - \frac{\lambda^2}{\tau^2}\Biggr)T = 0.
\end{equation}
   The solution is given by $ J_\lambda(\mu\tau)$
and/or $ N_\lambda(\mu\tau)$.  The
main difference between $J_\lambda(\mu\tau)$ and
$N_\lambda(\mu\tau)$ is their different behavior as $\mu\tau$ approaches $0$.
In the limit of a boost-invariance,
 $\lambda\rightarrow 0$, $J_\lambda (\mu\tau)$
approaches unity
while $N_\lambda(\mu\tau) \rightarrow \ln (\mu\tau)$.  If there is no
transverse momentum, the latter approaches the uniform solution  ($a = 0$,
$\kappa = b$) described in Sec.\ 2,
while the former coincides with the lowest energy
solution given in  (21).
   Putting all together,
one obtains the uniform solution in a complete form:
\begin{eqnarray}
  \theta(x){\bf n}_0 & = & \Bigl(a\:J_\lambda(\mu\tau) +
                      b\:N_\lambda(\mu\tau)\Bigr)\nonumber \\
             &   & \times \Bigl(A\:\cosh\lambda\eta +
                      B\:\sinh\lambda\eta\Bigr)\nonumber \\
             &   & \times J_m(\mu\rho)(C_m \cos m\phi +
           D_m \sin m\phi)\,{\bf n}_0\;{\rm \Theta}(\tau^2-{\bf x}_\bot^2),
\end{eqnarray}
where one may superpose these solutions in $\lambda$, $\mu$, and $m$.

 It is much harder to solve directly for nonuniform solutions when there is no
boost-invariance constraint. Though some simple special solutions can be
obtained by luck, finding all nonuniform solutions is a formidable task.
In contrast, it is straightforward to perform $SU(2)\times SU(2)$ rotations
on the uniform solutions.  The rotation formulas  (23) and (24) are most
general and applicable to the boost-noninvariant case as well.  Therefore
the nonuniform solutions can be obtained by the chiral rotation from
the uniform one in (36).

   An important question is whether we exhaust all nonuniform
solutions by the
axial rotations from the  uniform solutions.
In other words, are there any
nonuniform solutions that cannot be rotated into
a uniform one?  If such a class of solutions exists, it
would have some topological quantum number like a soliton.  Note however that
the solutions of our interest are time-dependent, and that their energies and
actions are not necessarily finite.
Unless these topologically nontrivial solutions exist,
the uniform solution  (36) and the $SU(2)\times SU(2)$ rotations on it
exhaust all solutions.

%
\section{Pion Charge Distribution}
   It has been predicted that the pions decaying from a DCC will show
 a distinct
charge distribution when the charge ratio is plotted event by event in
$f= N_{\pi^0}/(N_{\pi^0}+N_{\pi^\pm})$.
The distribution
\begin{equation}
\frac{dP}{df} =
\frac{1}{2\sqrt{f}} \label{dpdf}
\end{equation}
 has been derived in two very different ways.
The first derivation
assumes that an isosinglet multipion state is created by the decay of a
DCC \cite{horn}.
All pions decaying from a given DCC are assumed to occupy an identical orbital
state that is determined by the spatial configuration of DCC.
The Bose statistics allows only one isosinglet $2N$-pion state:
\begin{equation}
   |2N\pi\rangle =
        (2a_{+}^\dagger a_{-}^\dagger - a_0^\dagger a_0^\dagger)^N|0\rangle,
\end{equation}
where $a_{\pm,0}^\dagger$ are the creation operators of the pions in the same
single orbital state.  Making a binomial expansion of the right-hand side at
large $N$, one obtains a simple rule
$dP/df = 1/(2\sqrt{f})$. It is later pointed out that
the relative phase between
$a_{+}^\dagger a_{-}^\dagger$ and $a_0^\dagger a_0^\dagger$ is inessential to
the final prediction of $dP/df$ \cite{kog}.

   The second derivation is based on a more
intuitive picture in classical field theory.  Assuming
that the isospin orientation is uniform in space-time and
that all isospin directions are equally probable, one
obtains $dP/d\Omega = 1/4\pi$, where $\Omega$ is the solid angle for
an isospin
direction in isospin space.  Since the $\pi^0$ fraction $f$ is
proportional to the square
of the third component of the pion field,
($f\propto\cos^2\beta$), one obtains again distribution (\ref{dpdf}).
In this derivation,
the interference effects are completely ignored.

   Let us examine whether or not
this prediction remains valid for the nonuniform DCC's.
In the first derivation, it is crucial that only one orbital state is
available for pions and therefore the isosinglet state is unique:  For
two pions, the isosinglet is nothing but $(2a_{+}^\dagger a_{-}^\dagger -
a_0^\dagger a_0^\dagger)|0\rangle$ by the Clebsch-Gordan coefficients. For four
pions, the
group theory alone would allow two isosinglets.  One is to
combine the ${\bf 0}_{2\pi}$ from
${\bf 1}_\pi\otimes{\bf 1}_\pi ={\bf 0}_{2\pi}
+ {\bf 1}_{2\pi}$ with the other
${\bf 0}_{2\pi}$ from ${\bf 1}_\pi\otimes{\bf 1}_\pi ={\bf 0}_{2\pi}
+{\bf 1}_{2\pi}$. The other is to  contract the ${\bf 1}_{2\pi}$ from
${\bf 1}_\pi\otimes{\bf 1}_\pi= {\bf 0}_{2\pi} + {\bf 1}_{2\pi}$ with the other
${\bf 1}_{2\pi}$ from ${\bf 1}_\pi\otimes{\bf 1}_\pi={\bf 0}_{2\pi}+
{\bf 1}_{2\pi}$. The
Bose statistics forbids ${\bf 1}_{2\pi}$ for two identical pions
in the same orbital state so that only ${\bf 0}_{2\pi}\otimes {\bf 0}_{2\pi}
|0\rangle$ = $(2a_{+}^\dagger a_{-}^\dagger -a_0^\dagger a_0^\dagger)^2|0
\rangle$ is allowed. This argument goes through for any $2N$, leading to the
$|2N\pi\rangle$ in  (38).  If there are more than one orbital states
 available,
the four-pion singlet state would generally take the form
\begin{equation}
  |4\pi\rangle = \Bigl(A({\bf 0}_{2\pi}\otimes{\bf 0}_{2\pi})+B({\bf 1}_{2\pi}
                    \otimes{\bf 1}_{2\pi})\Bigr)|0\rangle,
\end{equation}
where the coefficients $A$ and $B$ are dynamics-dependent.  There are
increasingly many more isosinglets for $6\pi$'s, $8\pi$'s {\it etc}, as
$N$ goes up.
In the above example, the $A$-type term and the $B$-type term give
quite different pion compositions: there is $\pi^0\pi^0\pi^0\pi^0$ in the
$A$-type
term, but no $\pi^0\pi^0\pi^0\pi^0$ in the $B$-type term.
In order to obtain the distribution $dP/df = 1/(2 \sqrt{f})$, there must be
only
the $A$-type term and nothing else in the $2N\pi$ state ($N\rightarrow\infty$).

One can construct
explicitly the $4\pi$ state when the isovector field ${\bf n}(x)$ is nonuniform
in space-time.
Let us parametrize the direction of ${\bf n}(x)$ by the azimuthal
and polar angles $\alpha(x)$ and $\beta(x)$ with respect to the isospin z-axis.
To simplify our computation a little,
we consider as an example a DCC whose isospin is nonuniform only in the polar
direction $\beta$, but not in the azimuthal direction by choosing
 $\alpha
= 0$.  We shall use the representation
$\mbox{\boldmath $\pi$}(x)= f_{\pi}{\bf n}(x)\sin
\theta(x)$ instead of $\pi(x) = f_{\pi}{\bf n}(x)\theta(x)$ for the
following discussion
since the former automatically incorporates the periodicity of
$\Sigma(x)$ in $\theta(x)\rightarrow\theta(x)\pm 2\pi$.
The Cartesian isospin components of the pion field are
\begin{eqnarray}
 \pi_1 & =  &  f_{\pi}\sin\theta(x)\;\sin\beta(x),\nonumber\\
 \pi_2 & =  &  0, \nonumber \\
 \pi_3 & =  &  f_{\pi}\sin\theta(x)\;\cos\beta(x).
\end{eqnarray}
The DCC state is described by the quantum
coherent state,  up to an overall normalization
\begin{equation}
  |DCC(\theta,\beta)\rangle = \exp\Bigl(a_1^\dagger(s_\theta s_\beta)
                 + a_3^\dagger(s_\theta c_\beta)\Bigr)|0\rangle,
\end{equation}
where
\begin{eqnarray}
      a_1^\dagger(s_\theta s_\beta) & = &
                        \int\sqrt{2|{\bf k}|}\phi_{ss}
                      ({\bf k})a_1^\dagger({\bf k}) d^3{\bf k},\nonumber \\
      a_3^\dagger(s_\theta c_\beta) & = &
                        \int\sqrt{2|{\bf k}|}\phi_{sc}
                      ({\bf k})a_3^\dagger({\bf k}) d^3{\bf k},
\end{eqnarray}
with $\phi_{ss}$ and $\phi_{sc}$ being the three-dimensional Fourier transforms
of $f_{\pi}\sin\theta\sin\beta$ and $f_{\pi}\sin\theta\cos\beta$
respectively. Unlike $a_i^\dagger({\bf k})$, the operators
$a_1^\dagger(s_\theta s_\beta)$
and $a_3^\dagger(s_\theta c_\beta)$ are not canonically normalized, but the
normalization is irrelevant to the isospin structure.
The $|N\pi\rangle$ projection of the DCC state is
\begin{equation}
   |N\pi(\theta(x)\,\beta(x))\rangle
 =\frac{1}{N!}\Bigl(a_1^\dagger(s_\theta\,s_\beta) + a_3^\dagger
  (s_\theta\,c_\beta)\Bigr)^N|0\rangle.
\end{equation}
Under the assumption that the DCC's
appear in the intermediate state with $I=0$ and the
production processes conserve
isospin,  if one DCC can be produced, all other
DCC's that are related to it by the isospin rotations can be produced with
an equal probability.  The isosinglet DCC state can be constructed from
the state in  (41) by integrating out
 the Euler angles over the entire isospin
space.

   The $4\pi$ state of an isosinglet DCC is obtained by averaging
$|4\pi(\theta(x)\, \beta(x))\rangle$ over isospin space.  The
computation is
straightforward though a little tedious.  Up to an overall normalization,
the result is
\begin{equation}
 |4\pi(I=0)\rangle =
        \Biggl(\Bigl(\Bigl|{\bf a}^\dagger(s_\theta c_\beta)\Bigr|^2
               +\Bigl|{\bf a}^\dagger(s_\theta s_\beta)\Bigr|^2\Bigr)^2
              -4\Bigl|{\bf a}^\dagger(s_\theta c_\beta)\times
              {\bf a}^\dagger(s_\theta s_\beta)\Bigr|^2\Biggr)|0\rangle,
\end{equation}
where $|{\bf a}^\dagger|^2 = 2a_{+}^\dagger a_{-}^\dagger - a_0^\dagger
a_0^\dagger$.
The first and second terms in the right-hand side are the $A$-type terms
in  (39),
while the last term is the $B$-type term.
As we anticipate, the isosinglet $4\pi$
state of the nonuniform DCC is no longer of the form postulated in  (38).
For a uniform DCC, that is, $\beta(x)\rightarrow\,{\rm constant}$,
${\bf a}^\dagger(s_\theta c_\beta)$ and ${\bf a}^\dagger(s_\theta s_\beta)$
are identical up to a factor
(in the 1+1 boost-invariant solution in Sec.\ 2,
$\beta(x)$ is so defined that $\beta(x)\rightarrow 0$, namely
${\bf a}^\dagger\rightarrow 0$, in the uniform limit).
  Therefore, the $B$-type term
cannot exist for the uniform DCC's.
Our construction of the isosinglet $4\pi$ state and the existence of the
$B$-type terms
cast a serious doubt on the distribution for the nonuniform DCC's.

Alternatively,  let us study the problem by assuming that
$|{\rm DCC}(\theta(x),\beta(x))\rangle$ with
 different $\theta(x)$ and $\beta(x)$
do not have the quantum interference with each other.
It is in accordance with
the classical field picture. For a large number of pions, ignoring the
interference may be justified.
The momentum spectrum
of pion quanta ($i=1,2,3$) decaying from a classical field is given \cite{hen}
\begin{equation}
  (2\pi)^3\frac{dN_i}{d^3{\bf k}}
       = \frac{|\tilde{\rho}_i({\bf k},|{\bf k}|)|^2}{2|{\bf k}|},
        \label{dn}
\end{equation}
where $\tilde{\rho}_i({\bf k},|{\bf k}|)$
is the four-dimensional on-mass-shell
Fourier transform of the pion source function $\rho_i({\bf x},t)$ defined by
$\Box \mbox{\boldmath $\pi$}(x)=\mbox{\boldmath $\rho$}(x)$:
\begin{equation}
 \tilde{\rho}_i({\bf k}, |{\bf k}|) = \int \rho_i({\bf x}, t)\,
     e^{-i{\bf k}\cdot{\bf x} +i|{\bf k}|t}d{\bf x}dt.
\end{equation}
It is convenient to perform the space-time integral
using variables $\tau,\;\eta$ and
${\bf x}_\bot$ for which $d{\bf x}dt = \tau d\tau d\eta d{\bf x}_\bot$,
and
\begin{equation}
   E= |{\bf k}|, \;\;\; y =\frac{1}{2}{\rm ln}
           \Biggl(\frac{E+k_{\|}}{E-k_{\|}}\Biggr),
           \;\;\;{\bf k}_\bot ,\label{m}
\end{equation}
for the momentum variables. (\ref{dn}) becomes
\begin{equation}
  (2\pi)^3\frac{dN_i}{dy\,d^2{\bf k}_\bot}
  = \Bigl|\int \rho_i(\tau,\,\eta,\,{\bf x}_\bot)\;e^{i|{\bf k}_\bot|\tau
  \,\cosh(\eta - y) -i{\bf k}_\bot\cdot{\bf x}_\bot }
  \tau\,d\tau\,d\eta\,d^2{\bf x}_\bot\Bigr|^2.
\end{equation}
If a DCC is boost-invariant along the collision axis,
$\rho(\tau,\eta,{\bf x}_\bot)$ does not depend on $\eta$.
In this case, $\eta$ is integrated out and the energy spectrum
$dN_i/dy\,d^2{\bf k}_\bot$ is independent of the rapidity variable $y$,
as it is well known.

Let us look into the isospin structure of the Fourier transform of the source
that enters the right-hand side of  (48).  After the space-time
integration is performed, the isovector
$\tilde{\mbox{\boldmath $\rho$}}$ is generally
of the form
\begin{equation}
   \tilde{\mbox{\boldmath $\rho$}}({\bf k}, |{\bf k}|) =
F(y,{\bf k}_\bot)\mbox{\boldmath $e$}({\bf k})\; ,
\end{equation}
where $F(y,{\bf k}_\bot)$ is an isoscalar, Lorentz-scalar function of ${\bf k}$
and of whatever parameters that
characterize a DCC; $\mbox{\boldmath $e$}({\bf k})$ is a unit vector
in isospin space.
The pion spectrum is simply
\begin{equation}
  (2\pi)^3\frac{dN_i}{dy\,d^2{\bf k}_\bot}
  = |F(y,{\bf k}_\bot) e_i({\bf k})|^2.
\end{equation}
For each {\em fixed} ${\bf k}$, one may
repeat the classical field derivation
and reproduce
\begin{equation}
\frac{dP}{df({\bf k})} = \frac{1}{2\sqrt{f({\bf k})}}.\label{dpdfk}
\end{equation}
However,
it is clear that the pion number $N_i$ no
longer obeys distribution (\ref{dpdf}) when the momentum ${\bf k}$ is
integrated over.
To illustrate this point, consider a toy DCC for
which $\mbox{\boldmath $e$}({\bf k})$
points to one direction for a half of the range of
rapidity $y$ and to another direction perpendicular to it for the other half
of $y$.  Such a DCC is not one of the solutions that we have obtained, but it
serves to make a point.
Since there is no way to align the two ${\bf e}({\bf k})$'s to the same
direction by isospin rotation, there are no DCC's in this isospin family
that emit only $\pi^0$, even though all directions are equally probable
in isospin space.  For a family of nonuniform DCC's, $dP/df$
is zero at $f=0$ ({\em Centauro}) and at
$f=1$({\em anti-Centauro}), and tends to bulge in the central region of $f$,
unlike that for a family of uniform DCC's.  Only if the uniform
DCC's dominate over the nonuniform ones, can distribution (\ref{dpdf}) hold
approximately.
The abundance of the uniform DCC's has a measure zero
relative to that of the
 nonuniform
DCC's in the phase space of the rotation angles.  Unless the production of
the
nonuniform DCC's by the initial hadrons is
strongly suppressed for some dynamical
reason, $dP/df= 1/(2\sqrt{f})$ cannot hold even approximately.
The spectacular
 Centauro and anti-Centauro events will be far rarer than
our naive expectation based on the uniform DCC's.
However, there may be a chance to observe  distribution (\ref{dpdfk})
by selecting pions of the same $y$ and ${\bf k}_\bot$ within
small uncertainties.

A special case is  a boost-invariant DCC in $1+1$ dimensions
where $\mbox{\boldmath $\rho$}(x)$ is $\eta$-independent so that
$\tilde{\mbox{\boldmath $\rho$}}({\bf k})$
is $y$-independent.
In this case, $\mbox{\boldmath $e$}({\bf k})$
becomes a constant vector
independent of ${\bf k}$ and distribution (\ref{dpdf}) follows even for the
nonuniform DCC's. One the other hand,
for the $4\pi$ state that we studied in this Section,
we see nothing special about the boost-invariant nonuniform case
with ${\bf k}_\bot = 0$
from the general boost-noninvariant case. Do two derivations
contradict with each other?
It is difficult to make a connection between the two
arguments.  In analyzing the $|N\pi(I=0)\rangle$ state, the
interference between
different DCC's is essential while
in the classical field analysis, each DCC state is not an
eigenstate of isospin and the interference from different DCC's
is completely
discarded.  Though the both methods have led to the same $dP/df$ distribution,
it is not clear how much similar or mutually compatible their physical
pictures are. With this unsolved uncertainty,
we state our conclusion in
a less assertive way: if we follow the isospin analysis of $|2N\pi\rangle$,
we see no mechanism that leads to distribution (\ref{dpdf}) for
the nonuniform DCC's.
If we argue instead in
the classical field picture, the distribution does not hold
 except for the
boost-invariant DCC with zero ${\bf k}_\bot$. However,
distribution (\ref{dpdf})
should hold for pions which are selected from a small segment of rapidity $y$
and transverse momentum ${\bf k}_\bot$.

\section*{Acknowledgment}
We wish to thank J.D.\ Bjorken for giving us his
notes on his solutions and many useful discussions.
This work was supported in part by the Director, Office of Energy
Research, Office of High Energy and Nuclear Physics, Division of High Energy
Physics of the U.S. Department of Energy under contract DE-AC03-76SF00098 and
in part by the U.S. National Science Foundation under grant PHY-90-21139.
One of us (Z.H.) acknowledges the support from the
Natural Sciences and Engineering Research Council of Canada.

\newpage

\vskip 1cm
\noindent{\bf Figure caption}

\vskip 9pt

{\bf Figure 1}:
   The direction of ${\bf n}(\tau)$ relative to ${\bf a}$ and ${\bf b}$.
As $\tau$ varies, ${\bf n}(\tau)$ precesses in the plane
perpendicular to the vector ${\bf a}$. ${\bf c}={\bf b}\times {\bf a}$.
   For the uniform solutions,
${\bf n}$ stays in the direction of ${\bf b}$ for all $\tau$.


\newpage
\begin{thebibliography}{9}
\bibitem{horn}  D. Horn and R. Silver, Ann. Phys.(N.Y.) {\bf 66}, 509 (1971)
\bibitem{DCC} A. A. Anselm and M. G. Ryskin, Phys. Lett. B {\bf 266},
        482 (1991); J. -P. Blaizot and A. Krzywcki, Phys. Rev. D {\bf 46},
        246 (1992); J. D. Bjorken, K. L. Kowalski,
        and C. C. Taylor, SLAC preprint SLAC-PUB-6109.
\bibitem{rw} K. Rajagopal and F. Wilczek, Nucl. Phys. {\bf B399},
        395 (1992); {\it ibid} {\bf B404},
577 (1993).
\bibitem{blaiz}J.-P. Blaizot and A. Krzywicki, Phys. Rev. {\bf D46}, 246
(1992)
\bibitem{Bander} A. A. Anselm and M. Bander, Pis'ma Zh. Eksp. Teor. Fiz.
         {\bf 59}, 479 (1994) [JETP Lett. {\bf 59}, 503 (1994)].
\bibitem{kog} I.I. Kogan, Pis'ma Zh. Eksp. Teor. Fiz. {\bf 59}, 289 (1994)
          [JETP Lett. {\bf 59}, 307 (1994)]
\bibitem{hen}  E.M. Henley and W. Thirring, {\it Elementary Quantum
  Field Theory} (McGraw Hill, New York, 1962), Chapter 8-10
\end{thebibliography}
\end {document}
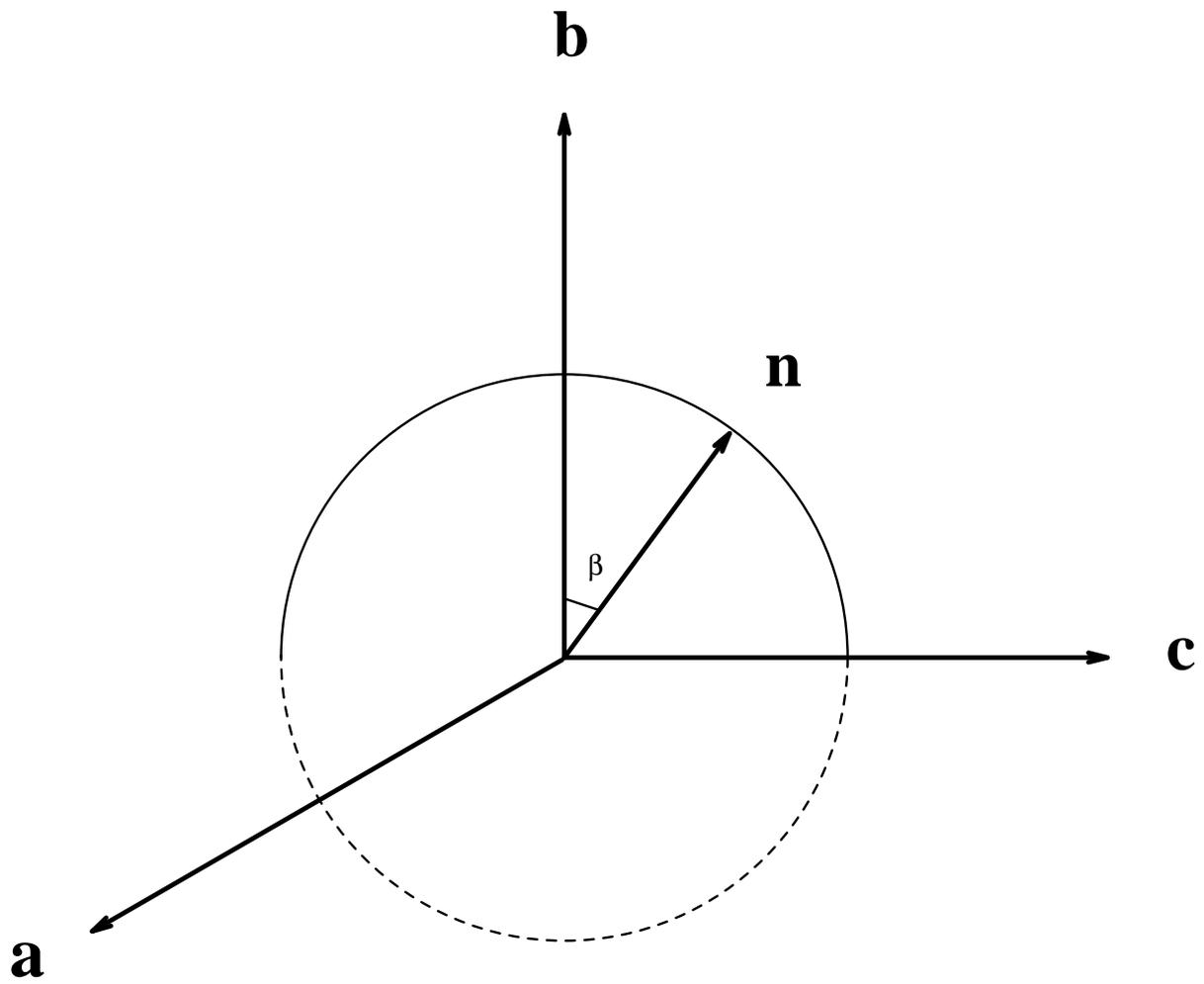
